\documentclass[aps,floatfix,twocolumn]{revtex4}
\usepackage{graphics,epsfig}
\usepackage{graphicx}
\usepackage{amsmath,amssymb}

\newcommand{\qC}{\mbox{$\mathcal{C}$}}

\begin{document}

\title{Transforming the Einstein static Universe into physically acceptable static fluid spheres II: A two - fold infinity of exact solutions}
\author{C\'edric Grenon, Pascal J. Elahi and Kayll Lake}
\affiliation{Department of Physics, Queen's University, Kingston,
Ontario, Canada, K7L 3N6 }
\date{\today}

\begin{abstract}
Following a solution generating technique introduced  recently by one of us, we transform the Einstein static Universe into a two - fold infinity class of physically acceptable exact perfect fluid solutions of Einstein's equations. Whereas the entire class of solutions can be considered as generalizations of the familiar Tolman IV solution, no member of the class can be written explicitly in isotropic coordinates. Further, except for a set of measure zero, no member of the class can be written explicitly in curvature coordinates either.
\end{abstract}

\maketitle
\section{Introduction}
Over the last decade or so the study of exact spherically symmetric perfect fluid solutions of Einstein's equations has been reinvigorated \cite{visser}. Recently, one of us \cite{lake} transformed the Einstein static Universe into an exact physically acceptable perfect fluid solution of Einstein's equations and examined the properties of the solution in familiar curvature coordinates. The emphasis there was on the importance of the integration constant that the generating technique introduced. Here we transform the Einstein static Universe into a two - fold infinity of solutions all of which may be considered generalizations of the familiar Tolman IV solution. (This solution, and its variants, is useful for the study of internal properties of neutron stars \cite{lat}.) It is shown that no member of the class can be written explicitly in isotropic coordinates and, except for a set of measure zero, no member can be written explicitly in curvature coordinates either.

When looking for an exact solution of Einstein's equations, the search most often starts from a familiar set of coordinates. The present paper serves to emphasize the important point that physically interesting solutions can be automatically excluded by this approach.
\section{Simplified Generating Technique}
Following \cite{lake} we consider spacetimes $\mathcal{O}$ that can be written in the form \cite{notation}
\begin{equation}
ds^2_{\mathcal{O}}=e^{2\chi(r)}\left(\frac{dr^2}{1-\frac{2M(r)}{r}}+r^2d\Omega^2-dt^2\right)\label{newform}
\end{equation}
where $d\Omega^2 \equiv d\theta^2+\sin(\theta)^2 d\phi^2$ and $M$ is constructed so as to make $\mathcal{O}$ a perfect fluid \cite{regular}. That is,
\begin{equation}
M=\frac{\int \!\tilde{b} ( r) {e^{\int \!\tilde{a} ( r)
{dr}}}{dr}+\mathcal{C}_{1}}{{e^{\int \!\tilde{a}(r) {dr}}}} \label{mass}
\end{equation}
where $^{'}\equiv\frac{d}{dr}$, $\mathcal{C}_{1}$ is a constant,
\begin{equation}
\tilde{a}\equiv\frac{4r^2(\chi^{''}-\chi^{'\;2})-3(1+2r\chi^{'})}{r(1+2r\chi^{'})}\label{newa}
\end{equation}
and
\begin{equation}
\tilde{b}\equiv\frac{2r(r(\chi^{''}-\chi^{'\;2})-\chi^{'})}{1+2r\chi^{'}}.\label{newb}
\end{equation}
It is important to note that $\mathcal{O}$ is not a conformal transformation of the seed metric given by the case $\chi=0$ (due to the restrictions on $M$). Moreover, $M$ is not the effective gravitational mass ($m$) of $\mathcal{O}$ \cite{mass}. The regularity conditions on $\chi$ are \cite{dellake}
\begin{equation}
|\chi(0)|<\infty,\;\;\;\;\chi^{'}(0)=0.\label{regular}
\end{equation}
Throughout we refer to the boundary (where the isotropic pressure vanishes and the solution matches onto vacuum) as $\Sigma$. Because of the form of (\ref{newform}), it is important to recall the following scaling property: If a spherically symmetric perfect fluid is represented by $ds^2$, then under the conformal transformation $ds^2 \rightarrow \delta^2 ds^2$, where $\delta$ is a constant, $p \rightarrow p/\delta^2$, $\rho \rightarrow \rho/\delta^2$ and $m \rightarrow \delta m$.

\section{Tolman IV}
\begin{figure}[ht]
\epsfig{file=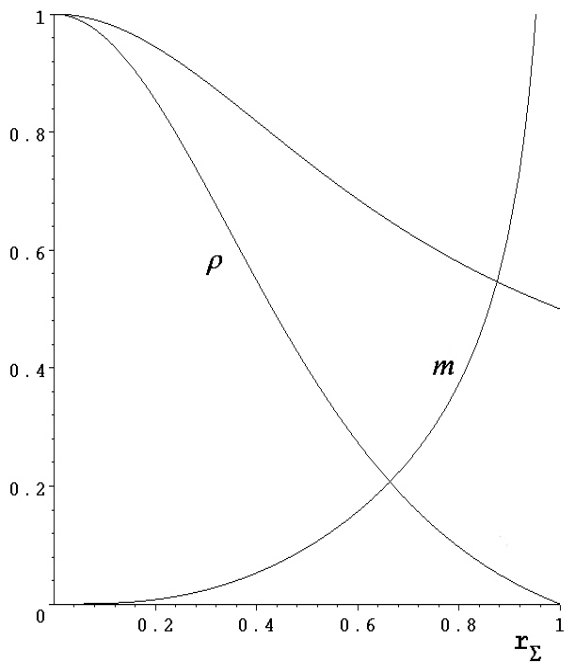,height=3.5in,width=3in,angle=0}
\caption{\label{tolmanproperties}Some boundary properties of the Tolman IV solution.
The abscissa is $r_{_{\Sigma}}$ defined by $p(r_{_{\Sigma}})=0$. For the top curve the ordinate is $-\mathcal{C}_{1}$. The next gives $\rho$ and so gives the density discontinuity at the boundary. The last curve gives the effective gravitational mass $m$, the total mass of each object.}
\end{figure}

\begin{figure}[ht]
\epsfig{file=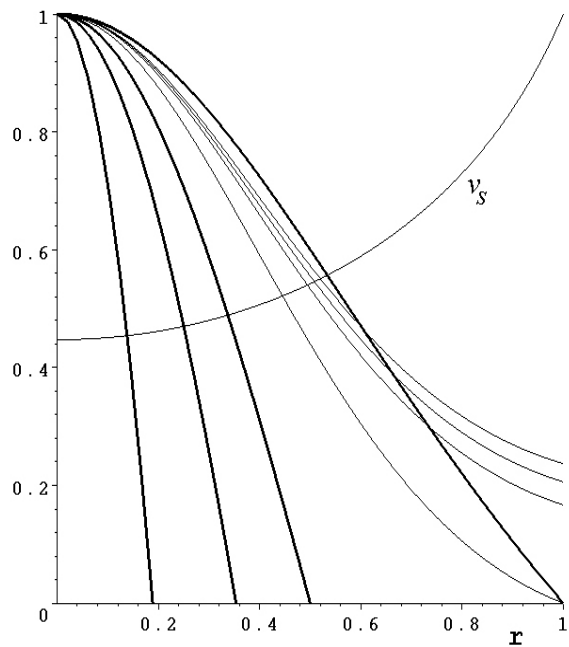,height=3.5in,width=3in,angle=0}
\caption{\label{tolmanpropertiesin}Some internal properties of the Tolman IV solution.
The horizontal curve gives the adiabatic sound speed for all $\qC_1$. The thick curves give $p/p(0)$ and the other curves give $\rho/\rho(0)$. The values of $\qC_1$, right to left for $p/p(0)$ and bottom up for $\rho/\rho(0)$, are $-0.5,\;-0.75,\;-0.85$ and $-0.95$.}
\end{figure}

As an introduction to what follows, we start with a consideration of the Tolman IV solution  \cite{dellake} \cite{tolman}. This solution is generated here in the following way: We choose
\begin{equation}\label{tolman}
e^{2\chi(r)}=\frac{\mathcal{C}_{2}^2}{1-\mathcal{C}_{3}^2r^2}.
\end{equation}
Since $\mathcal{C}_{2}$ can be absorbed into the scaling property mentioned above and $\mathcal{C}_{3}$ can be absorbed into the scale of $r$, there is no loss in generality in setting $\mathcal{C}_{2}=\mathcal{C}_{3} =1$. It now follows from (\ref{mass}) that
\begin{equation}\label{m1}
M=\frac{r^3(1-\mathcal{C}_1(1-r^2))}{1+r^2}.
\end{equation}
The choice (\ref{tolman}) obviously introduces a coordinate singularity at $r=1$. This is discussed below.
The isotropic pressure and energy density are given by
\begin{equation}\label{tolmanpressure}
8\pi p(r)=\frac{r^2(1+4\qC_1)+2(1+\qC_1)}{1+r^2}
\end{equation}
and
\begin{equation}\label{tolmandensity}
8\pi\rho(r)=-\frac{r^4(1+4\qC_1)+r^2(5+2\qC_1)+6\qC_1}{(1+r^2)^2}
\end{equation}
respectively. Bearing in mind the scale factor $\delta$, we note that $\qC_1=-4 \pi \rho(0)/3=4 \pi p(0)-1$. The adiabatic sound speed is independent of $\qC_1$ and is given by
\begin{equation}\label{tolmansound}
v_{s}^2=\frac{1+r^2}{5-3r^2}.
\end{equation}
Solving for $\qC_1$ at $\Sigma$ we obtain
\begin{equation}\label{boundarytolman}
\qC_1=-\frac{2+r^2}{2(1+2r^2)} \Big |_{_{\Sigma}}
\end{equation}
which shows that $-1<\qC_1\leq-1/2$ and it follows that both $p$ and $\rho$ are monotone decreasing. Some boundary properties of this solution are shown in  FIG.~\ref{tolmanproperties} and some internal properties are shown in FIG.~\ref{tolmanpropertiesin}. The case $\qC_1=-1/2$ does not correspond to a finite object and the associated total mass $m$ diverges as $r \rightarrow 1 \equiv \Sigma$. This case is also distinguished by the peculiarity that all polynomial invariants of the Riemann tensor vanish identically as  $r \rightarrow 1$.  The coordinate singularity at $r=1$ plays no role throughout the range of finite objects, $-1<\qC_1<-1/2$.

Under the coordinate transformation
\begin{equation}\label{tolmantrans}
\frac{r(\mathsf{r})}{\sqrt{1-r(\mathsf{r})^2}}=\mathsf{r}
\end{equation}
the solution appears in the more familiar form
\begin{equation}\label{TolmanIV}
ds^2=\frac{(1+2\mathsf{r}^2)d\mathsf{r}^2}{(1+(1+2\mathcal{C}_1)\mathsf{r}^2)(1+\mathsf{r}^2)}+\mathsf{r}^2d\Omega^2-(1+\mathsf{r}^2)dt^2.
\end{equation}

\section{Integer Solutions}
We now consider
\begin{equation}\label{ntolman}
e^{2\chi(r)}=\frac{\mathcal{C}_{2}^2}{(1-\mathcal{C}_{3}^2r^2)^n}
\end{equation}
where $n$ is and integer $>1$. As explained above, there is no loss in generality in setting $\mathcal{C}_{2}=\mathcal{C}_{3} =1$.
It now follows from (\ref{mass}) that
\begin{equation}\label{M}
M=\frac{(-1)^{n-2}r^3}{(1-r^2)^{n-2}\mathcal{P} }\left(\psi\; \mathcal{P}\; \mathcal{H} +\mathcal{C}_1\right)
\end{equation}
where
\begin{equation}\label{psi}
\psi \equiv -\frac{1}{8}\frac{(2n-1)^{3-n}}{3n-2}(-2)^n n^{n-2}(n-2),
\end{equation}
\begin{equation}\label{P}
\mathcal{P}\equiv\epsilon^\frac{3n-2}{2n-1},
\end{equation}
\begin{equation}\label{epsilon}
\epsilon \equiv (2n-1)r^2+1,
\end{equation}
and
\begin{equation}\label{H}
\mathcal{H}\equiv {}_2F_1\left(3-n,\frac{3n-2}{2n-1};\frac{5n-3}{2n-1};\frac{\epsilon}{2n}\right).
\end{equation}
Here $F$ is the standard hypergeometric function defined by
\begin{equation}
{}_2F_1(a,b;c;z)\equiv \sum^\infty_{k=0}\frac{(a)_k(b)_k}{(c)_k}\frac{z^k}{k!}
\end{equation}
where $(x)_k$ is the Pochhammer symbol
\begin{equation}
(x)_k\equiv \frac{\Gamma(x+k)}{\Gamma(x)}=x(x+1)\dots(x+k+1).
\end{equation}
All these solutions are physically distinct from, but bear some similarities to, the Tolman IV case $n=1$ \cite{lake2}.
The case $n=2$ is treated separately in the Appendix. For $n>2$ the effective gravitational mass is given by
\begin{equation}\label{effectivemass}
m={\frac {M( 2(n-1)^{2}{r}
^{4}+4(n-1) {r}^{2}+2 )-n(n-2){r
}^{5}-2n{r}^{3}}{ 2(1-{r}^{2})^{2+n/2}}}
\end{equation}
where $M$ is given by (\ref{M}).

Let us define the ``tenuity"
\begin{equation}\label{tenuity}
\alpha \equiv \sqrt{\frac{A}{4 \pi m^2}}\Big |_{_{\Sigma}}
\end{equation}
where $A$ signifies the area of the two-surface $t=const$ and $r=r|_{_{\Sigma}}$. Equivalently \cite{mass},
\begin{equation}\label{tenuityangle}
\alpha=\frac{2}{g_{\theta \theta}\;\mathcal{R}_{\theta \phi}^{\;\;\;\; \theta \phi}}\Big |_{_{\Sigma}}.
\end{equation}
We find
\begin{equation}\label{tenuityn}
\alpha=\frac{(3n-1)r^2+1}{nr^2}\Big |_{_{\Sigma}}
\end{equation}
for all $n$ and we note that $\alpha > 3$ for all finite objects in this class of solutions.
\subsection{Limiting solutions}

\begin{figure}[ht]
\epsfig{file=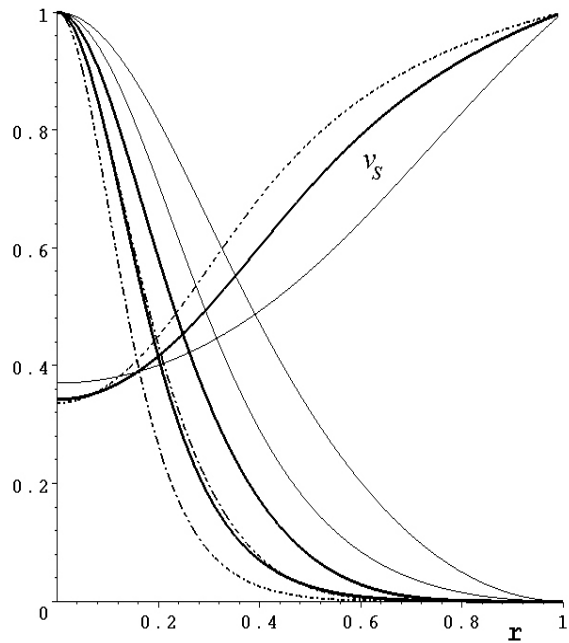,height=3.5in,width=3in,angle=0}
\caption{\label{global}Some internal properties of the global solutions. For the thin curves $n=2$, for the thick curves $n=5$ and for the dashdot curves $n=8$. The horizontal curves give the adiabatic sound speed. The vertical curves give $p/p(0)$ on top and $\rho/\rho(0)$ below.}
\end{figure}

As in the case of Tolman IV, for each $n>1$ there exists a global solution, characterized by a unique value of $\qC_1$. These do not correspond to a finite object and the total mass $m$ diverges as $r \rightarrow 1 \equiv \Sigma$ (as does the coordinate mass $M$ for $n>2$). These solutions share the distinction that all polynomial invariants of the Riemann tensor vanish identically as  $r \rightarrow 1$. For $n=2$, the critical value of $\qC_1$ is $2^{5/3}$. For $n>2$ the critical values are given by
\begin{equation}\label{c1special}
\qC_1 = -(\psi\; \mathcal{P}_{r=1}  \mathcal{H}_{r=1})
\end{equation}
where $\psi$ is given by (\ref{psi}), $\mathcal{P}_{r=1}$ and $\mathcal{H}_{r=1}$ by (\ref{P}) and (\ref{H}) respectively with $\epsilon=2n$. Some properties of these global solutions are shown in FIG.~\ref{global}.

\subsection{Even Integers}
\begin{figure}[ht]
\epsfig{file=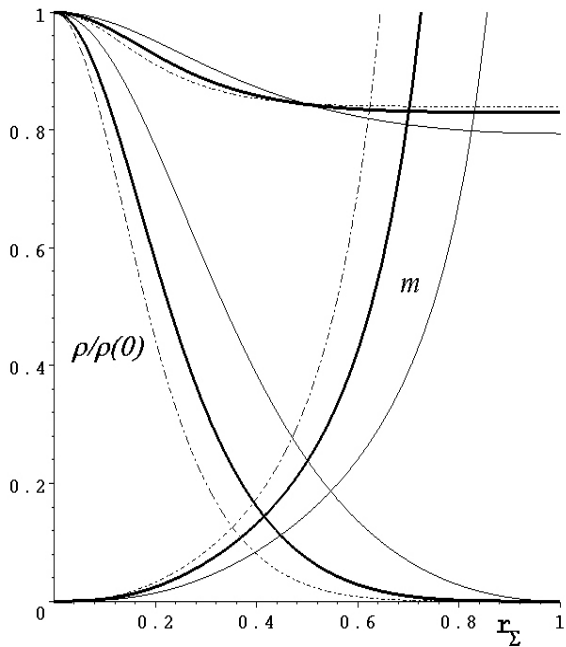,height=3.5in,width=3in,angle=0}
\caption{\label{evenproperties}Some boundary properties for even integers. For the thin curves $n=2$, for the thick curves $n=4$ and for the dashdot curves $n=6$.
The abscissa is $r_{_{\Sigma}}$ defined by $p(r_{_{\Sigma}})=0$. For the top curve the ordinate is $\mathcal{C}_{1}$ scaled by its maximum value at the center. The next gives $\rho/\rho(0)$ and so gives the scaled density discontinuity at the boundary. The last curve gives the effective gravitational mass $m$, the total mass of each object.}
\end{figure}

\begin{figure}[ht]
\epsfig{file=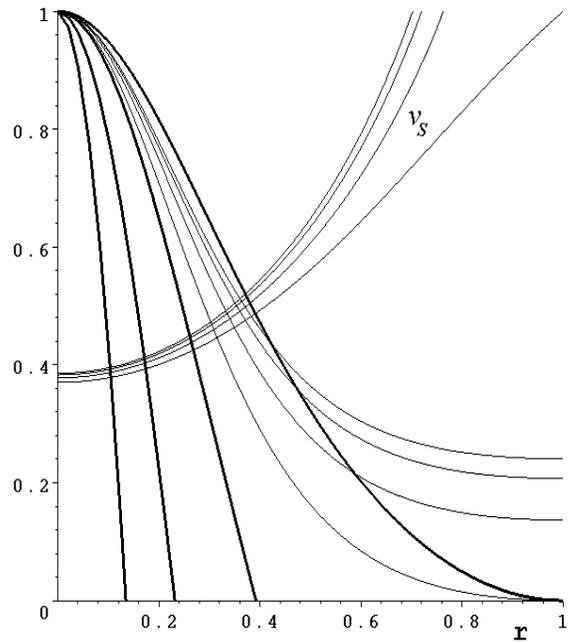,height=3.5in,width=3in,angle=0}
\caption{\label{evenpropertiesin}Some internal properties of the $n=2$ solution (explicit properties given in the Appendix).
The horizontal curves give the adiabatic sound speed. The thick curves give $p/p(0)$ and the other curves give $\rho/\rho(0)$. The values of $\qC_1$, right to left for $p/p(0)$ and bottom up for $\rho/\rho(0)$ and the sound speed, are $2^{5/3},\;3.5,\;3.75$ and $3.9$. Solutions with $n>2$ show qualitatively the same behaviour.}
\end{figure}

For $n$ even, all allowed values of $\qC_1$ are $>0$ and for $n>2$ the maximum allowed value of $\qC_1$ is obtained from the limit $r \rightarrow 0 \equiv \Sigma$ and the minimum allowed value of $\qC_1$ is obtained from the limit $r \rightarrow 1 \equiv \Sigma$ as given by (\ref{c1special}). For $n=2$, $2^{5/3}<\qC_1<4$ and for $n>2$
\begin{equation}\label{c1specialeven}
-(\psi\; \mathcal{P}_{r=1}  \mathcal{H}_{r=1})< \qC_1 < -(\psi\; \mathcal{H}_{r=0})+2n
\end{equation}
where $\mathcal{H}_{r=0}$ is given by (\ref{H}) with $\epsilon=1$. The coordinate singularity at $r=1$ plays no role throughout the range of finite objects. Bearing in mind the scale factor $\delta$, we find
\begin{equation}\label{meaningeven}
\qC_1=-4 \pi p(0)-(\psi\; \mathcal{H}_{r=0})+2n
\end{equation}
and note that
\begin{equation}\label{central}
p(0)=\frac{n}{4 \pi}-\frac{\rho(0)}{3}
\end{equation}
as in the case of Tolman IV ($n=1$).
Some boundary properties of the solutions for even $n$ are shown in  FIG.~\ref{evenproperties} and some internal properties are shown in FIG.~\ref{evenpropertiesin}.
\subsection{Odd Integers}
For $n$ odd, all allowed values of $\qC_1$ are $<0$ and the minimum allowed value of $\qC_1$ is obtained from the limit $r \rightarrow 0 \equiv \Sigma$ and the maximum allowed value of $\qC_1$ is obtained from the limit $r \rightarrow 1 \equiv \Sigma$ as given by (\ref{c1special}). For $n$ odd we have
\begin{equation}\label{c1specialodd}
-2n-(\psi\; \mathcal{H}_{r=0})< \qC_1 < -(\psi\; \mathcal{P}_{r=1}  \mathcal{H}_{r=1}).
\end{equation}
Again, the coordinate singularity at $r=1$ plays no role throughout the range of finite objects.  Again, bearing in mind the scale factor $\delta$, we find
\begin{equation}\label{meaninodd}
\qC_1= 4 \pi p(0)-(\psi\; \mathcal{H}_{r=0})-2n
\end{equation}
and (\ref{central}) also holds.
Boundary properties of the solutions for odd $n$ are qualitatively similar to FIG.~\ref{evenproperties} and their internal properties are qualitatively similar to FIG.~\ref{evenpropertiesin}.

\subsection{Coordinate Transformations}
In order to transform the foregoing solutions into explicit curvature coordinates, we must solve for $r(\mathsf{r})$ from the generalized form of (\ref{tolmantrans}),
\begin{equation}\label{curvature}
\frac{r(\mathsf{r})}{(1-r(\mathsf{r})^2)^{n/2}}=\mathsf{r}.
\end{equation}
Whereas this is possible in principle only for $n<5$, the only value of $n$ which we find gives a manageable metric is the Tolman IV case $n=1$.
In order to transform the foregoing solutions into explicit isotropic coordinates, we must solve for $r(\mathsf{r})$ from
\begin{equation}\label{curvaturet}
ln(\mathsf{r})=\int\frac{dr}{r \sqrt{1-2M(r)/r}}
\end{equation}
where $M(r)$ is given by (\ref{M}). We have found that this is not possible for any $n$.

\section{NON INTEGER SOLUTIONS}\label{fraction}
\begin{figure}[ht]
\epsfig{file=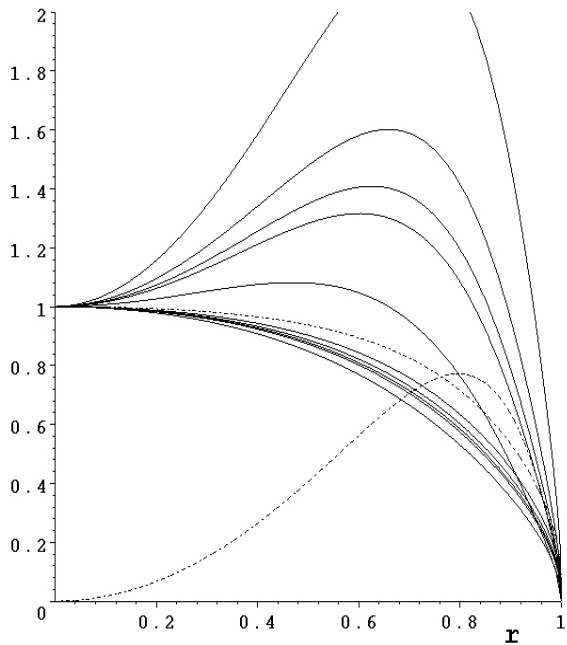,height=3.5in,width=3in,angle=0}
\caption{\label{nonintegerglobal}Some properties of the non integer global solutions. For $i=1$ (dashdot curves) $\rho$ and $p/p(0)$ are shown. The other cases show $\rho/\rho(0)$ and $p/p(0)$. In all cases $p/p(0)$ is monotone decreasing but $\rho/\rho(0)$ is not. The values of $i$ are $2,\;3,\;4,\;5$ and $100$ (essentially the same as $n=1/2$). These are read top down on both sets of curves.}
\end{figure}

Let us now consider (\ref{ntolman}) but with $n$ generated by
\begin{equation}
n=\frac{i}{1+2i}
\end{equation}
where $i$ is an integer $\geq 1$. This choice also admits global solutions but only with $\qC_1=0$. These solutions all have monotone decreasing $p$ but $\rho$ fails to be monotone (indeed $\rho(0)=0$ for $i=1$) and so these solutions have to be considered unphysical. Some properties of these solutions are shown in  FIG.~\ref{nonintegerglobal} for completeness. Non global solutions exist for $\qC_1\neq0$ but for these $\rho$ also fails to be monotone.

\section{Discussion}
The Einstein static Universe has been transformed into an explicit and exact two - fold infinity of physically acceptable perfect fluid solutions of Einstein's equations. The solutions are physically interesting because they are useful for the study of internal properties of neutron stars. The emphasis here is on the role of coordinates. Whereas the entire class of solutions can be considered as generalizations of the familiar Tolman IV solution, no member of the class can be written explicitly in isotropic coordinates and, except for a set of measure zero, no member of the class can be written explicitly in curvature coordinates either.

\begin{acknowledgments}
CG is supported by the Ontario Graduate Scholarship in Science and Technology and the Fonds qu\'eb\'ecois de recherche sur la nature et les technologies. PJE is supported by a scholarship from the National Science and Engineering
Research Council of Canada. KL is supported by a grant from the Natural Sciences and
Engineering Research Council of Canada. Portions of this work were made possible by use of \emph{GRTensorII} \cite{grt}.
\end{acknowledgments}

\bigskip

\appendix*
\section{$n=2$}
For $n=2$ we find
\begin{equation}\label{n2apmm}
M={\frac {\qC_1{r}^{3}}{( 1+3{r}^{2}) ^{4/3}}},
\end{equation}
\begin{equation}\label{n2apm}
m={\frac {{r}^{3}(2(1+3{r}^{2})^{4/3}-\qC_1(1+{r}^{2})^{2}) }{(1-r^2)^{
3}(1+3{r}^{2})^{4/3}}},
\end{equation}
\begin{equation}\label{n2app}
8\pi p=2{\frac {2(1+3{r}^{2})^{4/3}(2+{r}^{2}
)-\qC_1(1+5{r}^{2})(1+{r}^{2}
) }{(1+ 3{r}^{2})^{4/3}}},
\end{equation}
so that
\begin{equation}\label{n2app0}
\qC_1=-4 \pi p(0)+4
\end{equation}
and finally
\begin{equation}\label{n2aprho}
8 \pi \rho=2{\frac {\qC_1
 ( 5{r}^{6}+29{r}^{4}+11{r}^{2}+3)-6(1+ 3{r}^{2})^{7/3}}{(1+ 3{r}^
{2}) ^{7/3}}}
\end{equation}
so that
\begin{equation}\label{n2aprho0}
\qC_1=\frac{4}{3} \pi \rho(0)+2.
\end{equation}
The square of the adiabatic sound speed is given by
\begin{equation}\label{n2apvs}
{\frac {\left(\qC_1( 5{r}^{4}+2{r}^{2}+1) -
 (1+3{r}^{2})^{7/3}\right)(1+3{r}^{2}
) }{\qC_1( 1-{r}^{2})( 5{r}^{4}-2
\,{r}^{2}+5) }}.
\end{equation}


\begin{thebibliography}{}
\bibitem{notation}We use geometrical units and usually designate functional dependence only on the first appearance of a function. Throughout $\rho$ desinates the energy density and $p$ the isotropic pressure.
\bibitem{regular}The solutions we are interested in must satisfy (i) isotropy of the
pressure (otherwise any metric is a ``solution"), (ii) regularity
of associated invariants at the origin, (iii) positivity of the pressure and energy
density, (iv) vanishing of the pressure at a finite
boundary, (v) monotone decrease of the pressure and energy density to the
boundary and (vi) subluminal adiabatic sound speed.
\end{thebibliography}
\end{document}